\documentclass[pdflatex,sn-mathphys-num]{sn-jnl}
\usepackage{graphicx}%
\usepackage{multirow}%
\usepackage{amsmath,amssymb,amsfonts}%
\usepackage{amsthm}%
\usepackage{mathrsfs}%
\usepackage[title]{appendix}%
\usepackage{xcolor}%
\usepackage{textcomp}%
\usepackage{manyfoot}%
\usepackage{algorithm}%
\usepackage{algorithmicx}%
\usepackage{algpseudocode}%
\usepackage{listings}%
\usepackage[nomain,nopostdot,acronym]{glossaries}
\usepackage{mathtools}
\usepackage{subfig}
\usepackage{bbm}
\usepackage{bm}
\usepackage{tabularx}
\usepackage{adjustbox}
\usepackage{caption}

\usepackage{tikz}
\usetikzlibrary{external}
\usepackage{pgfplots}
\pgfplotsset{compat=1.18}    
\usepgfplotslibrary{groupplots}   
\usetikzlibrary{shapes.arrows}
\usetikzlibrary{plotmarks}
\usetikzlibrary{patterns}
\usetikzlibrary{arrows.meta}
\usetikzlibrary{positioning}
\usetikzlibrary{spy}

\unnumbered

\theoremstyle{thmstyleone}%
\theoremstyle{thmstyletwo}%

\theoremstyle{thmstylethree}%

\begin{document}
\glsdisablehyper 

\newacronym{roc}{ROC}{Receiver-operating characteristic}
\newacronym{cdf}{CDF}{Cumulative distribution function}
\newacronym{auc}{AUC}{Area Under the curve}
\newacronym{std}{STD}{Standard Deviation}
\newacronym{LEO}{LEO}{Low Earth Orbit}
\newacronym{GEO}{GEO}{Geostationary Earth Orbit}
\newacronym{UL}{UL}{Uplink}
\newacronym{DL}{DL}{Downlink}
\newacronym{PoP}{PoP}{Point of Presence}
\newacronym{RTT}{RTT}{Round-Trip Time}
\newacronym{tp}{TP}{True Positive}
\newacronym{fp}{FP}{False Positive}
\newacronym{fn}{FN}{False Negative}
\newacronym{tn}{TN}{True Negative}
\newacronym{QoE}{QoE}{Quality of Experience}
\newacronym{PR}{PR}{Precision-Recall}
\newacronym{udp}{UDP}{User Datagram Protocol}
\newacronym{irtt}{IRTT}{Isochronous Round-Trip Tester}
\newacronym{AUPRC}{AUPRC}{Area Under the PR-Curve}
\newacronym{DSA}{DSA}{Discounted Service Availability}
\newacronym{FPR}{FPR}{False Positive Rate}
\newacronym{IRIS2}{IRIS2}{Interconnectivity and Security by Satellite}
\newacronym{E2E}{E2E}{End-To-End}
\newacronym{SLA}{SLA}{Service Level Agreement}
\newacronym{ICMP}{ICMP}{Internet Control Message Protocol}
\newacronym{QoS}{QoS}{Quality of Service}
\newacronym{MSE}{MSE}{Mean Squared error}
\newacronym{NTP}{NTP}{Network Time Protocol}
\newacronym{GMM}{GMM}{Gaussian Mixture Model}
\newacronym{GPD}{GPD}{Generalized Pareto Distribution}
\newacronym{EVT}{EVT}{Extreme Value Theory}
\newacronym{NTN}{NTN}{Non Terrestrial Network}

\title{Statistical Characterization and Prediction of E2E Latency over LEO Satellite Networks}

\author[1]{\fnm{Andreas} \sur{Casparsen}}\email{aca@es.aau.dk}

\author[1]{\fnm{Jonas Ellegaard} \sur{Jakobsen}}\email{jej@es.aau.dk}

\author[1]{\fnm{Jimmy Jessen} \sur{Nielsen}}\email{jjn@es.aau.dk}

\author[1]{\fnm{Petar} \sur{Popovski}}\email{petarp@es.aau.dk}

\author[1]{\fnm{Israel Leyva} \sur{Mayorga}}\email{ilm@es.aau.dk}

\affil[1]{Department of Electronic Systems, Aalborg University, Aalborg, Denmark}

\abstract{\gls{LEO} satellite networks are emerging as an essential communication infrastructure, with standardized 5G-based non-terrestrial networks and their integration with terrestrial systems envisioned as a key feature of 6G. However, current \gls{LEO} systems still exhibit significant latency variations, limiting their suitability for latency-sensitive services.
We present a detailed statistical analysis of end-to-end latency based on 500Hz experimental bidirectional one-way measurements and introduce a segmentation of the deterministic 15-second periodic behavior observed in Starlink. We characterize handover-induced boundary regions that produce latency spikes lasting approximately 140 ms at the beginning and 75 ms at the end of each cycle, followed by a stable intra-period regime, enabling accurate short-term prediction.
This analysis shows that latency prediction based on long-term statistics leads to pessimistic estimates. In contrast, by exploiting the periodic structure, isolating boundary regions, and applying lightweight parametric and non-parametric models to intra-period latency distributions, we achieve 99th-percentile latency prediction errors below 50 ms. Furthermore, period-level latency prediction and classification enable adaptive transmission strategies by identifying upcoming periods where application latency requirements cannot be satisfied, necessitating the use of alternative systems.
\glsresetall}


\maketitle 
\section{Introduction}
Ubiquitous satellite-based internet access is rapidly expanding through \gls{LEO} constellations such as SpaceX’s Starlink, Iridium, and OneWeb, with additional systems under development, including Amazon LEO and the EU’s \gls{IRIS2}. Compared to traditional \gls{GEO} systems, \gls{LEO} networks offer substantially lower latency, broader coverage, and more flexible routing. While they cannot match the capacity of terrestrial fiber and cellular infrastructure, they provide valuable advantages in coverage and independence from ground networks. In some scenarios, such as rural regions, satellite links may even offer a more stable latency than cellular connectivity, albeit higher \cite{lopez2023connecting}.
Unlike terrestrial infrastructure, \gls{LEO} constellations are inherently dynamic: orbital motion induces predictable yet disruptive variations in the network geometry that affect latency.
This predictability concerns deterministic topology changes—not the underlying tropospheric or shadowing-induced radio fading, which remains random.
While such geometric regularity can benefit network management through scheduled reconfigurations, it also degrades \gls{QoS} for latency-sensitive applications by introducing variations from handovers, ground-station transitions, and dynamic routing \cite{zhao2024low}. These events not only increase latency but can also introduce transient packet-loss, throughput reductions, which together contribute to the overall performance variability observed.
Long-term latency studies confirm this effect, consistently reporting higher variance and frequent outliers in Starlink compared to terrestrial networks across multiple regions \cite{ma2023network}.

Previous studies have observed the effect on latency and throughput of such reconfigurations in the Starlink network. Empirical studies have consistently identified a~15-second periodicity in Starlink latency and throughput, attributed to deterministic handovers between satellites or ground stations \cite{mohan2024multifaceted, pan2024measuring, garcia2024fine}. These works observe that latency remains relatively stable within each period but shifts significantly between periods, often accompanied by drops in throughput.
Related observations of latency volatility caused by mobility and handover mechanisms are reported in \cite{inflation}, which characterizes latency inflation in mobile and satellite networks and identifies handovers as a dominant source of transient delay spikes. However, such studies do not resolve the fine-grained temporal structure or duration of these spikes within operational LEO systems.
To address the uncertainty in satellite networks, \cite{tiwari2023t3p} introduces a middleware that predicts short-term latency and throughput at 1 Hz, helping applications anticipate degradation. However, its coarse 1 Hz sampling obscures fine-grained variations during handovers and does not fully capture the short-term effects.
For example, \cite{rojas2025latency} studies the latency trade-off between space-based and terrestrial cloud computing, identifying distance-dependent regimes where satellite or terrestrial processing is preferable, while \cite{van2019low} applies geometry-based modeling to decompose propagation and routing latency in LEO constellations.
These studies provide valuable insights into long-term latency behavior and architectural trade-offs, but they do not capture short-term latency dynamics at fine temporal resolution, nor do they address repeatable intra-period structures observed in operational LEO systems.

Beyond latency characterization, works also investigate how latency knowledge can be exploited at the system or application level. For instance, \cite{liu2025vivisecting} leverages throughput and latency prediction to improve video streaming quality over Starlink via application-layer adaptation, while \cite{zhu2024latency} proposes intelligent scheduling mechanisms that use network-state information to reduce latency under congestion in large-scale satellite networks.
Approaches to mitigate latency and throughput fluctuations in \gls{LEO} satellite systems include multi-connectivity \cite{lopez202212} and application-level adaptations \cite{zhao2024low}. Multi-connectivity relies on the use of multiple interfaces via packet duplication or interface selection to increase reliability, but it is inefficient when the statistical properties of the individual interfaces are unknown. Similarly, application-level adaptations rely on predicting changes in the statistics of the network performance to adapt transmissions. For instance, a cloud gaming platform might proactively pre-buffer assets and data before a predicted high-latency interval, reducing the risk of rendering artifacts. Similarly, an edge computing system could delay non-critical transmissions during volatile windows to preserve responsiveness. While such predictions can be made using AI methods \cite{zhao2024low}, a statistical analysis might provide an equally accurate and more explainable solution.
Short-term latency prediction is particularly relevant for certain application domains. Latency-sensitive and safety-critical services - such as remote operation, real-time telemetry, cyber-physical control loops, and systems employing multi-connectivity - benefit from advance knowledge of whether an upcoming interval will satisfy a latency requirement. Such information enables concrete actions, including adjusting the operation mode (e.g., slowing or halting vehicles), deferring transmissions, or selecting an alternative interface, motivating the need for methods that characterize and predict latency over short time horizons in \gls{LEO} systems.

Furthermore, research indicates that novel solutions are required as traditional protocols for data transport on terrestrial networks find limited potential when applied to satellite constellations. 
\cite{garcia2023multi} highlights this by showing TCP under-utilizes Starlink bandwidth due to mismatched expectations, reporting a 46\% reduction in throughput efficiency, underscoring a fundamental mismatch between TCP’s congestion-control assumptions and satellite link dynamics. Meanwhile, \cite{lai2022spacertc} suggests a novel utilization of satellites in broader networks that enhances application performance over cloud. 
This line of research suggests that major reconsiderations for how protocols and systems are designed are necessary when a satellite link is incorporated. Traditional solutions to ensure high and stable \gls{QoE} are insufficient for fully utilizing the functionality and features of non-terrestrial communication.  
This further underscores the need for accurate short-term characterization and prediction of latency dynamics in satellite networks. Traditional solutions to ensure high and stable \gls{QoE} are insufficient for fully exploiting the functionality and features of non-terrestrial communication.

While prior studies have identified the existence of Starlink’s 15-second periodicity, they have not resolved or modeled the fine-grained intra-period latency structure due to coarser sampling rates or differing measurement objectives. In contrast, our high-rate measurements reveal the detailed boundary regions and intra-period dynamics that enable accurate short-term latency prediction.
Motivated by the need to reconsider how to utilize satellite communication more efficiently, we propose a statistical framework for \gls{E2E} latency characterization and prediction in \gls{LEO} satellite systems, which operates on high-rate latency measurements to segment 15-second periods, isolate handover-induced latency spikes, and model intra-period latency distributions.
Specifically, this framework leverages the known 15-second periodic structure observed in Starlink \cite{mohan2024multifaceted, pan2024measuring, garcia2024fine} to predict short- and long-term latency statistics within each period. We perform a thorough statistical analysis of the latency of Internet connectivity through \gls{LEO} constellations with Starlink as a case study and characterize the deterministic and stochastic temporal dynamics of latency. The proposed latency characterization and classification approach can be utilized by applications to predict the \gls{QoS} and service availability up to 15 seconds in advance, and to adapt the application and transmission parameters according to the predicted latency conditions, thereby maximizing \gls{QoE} and guaranteeing \gls{SLA}. The main technical contributions of this work are as follows. 
\begin{itemize}
\item We design a provider-agnostic testbed at Aalborg University for accurate \gls{E2E} latency measurements across \gls{UL}, \gls{DL}, and round-trip paths.
\item Using multi-month Starlink traces, we provide a fine-grained statistical characterization of its deterministic 15-second latency structure, distinguishing boundary-induced spikes from the stable intra-period core and comparing intra-period latency distributions and high quantiles for \gls{UL}, \gls{DL}, and \gls{RTT}.
\item We evaluate parametric and non-parametric methods for predicting latency quantiles within each cycle, enabling short-term application adaptation.
\item We introduce a statistical framework for predicting \gls{SLA}-compliant service availability under latency and reliability constraints up to 15 seconds ahead.
\end{itemize}

\begin{figure}
    \centering
    \includegraphics[width=0.8\linewidth]{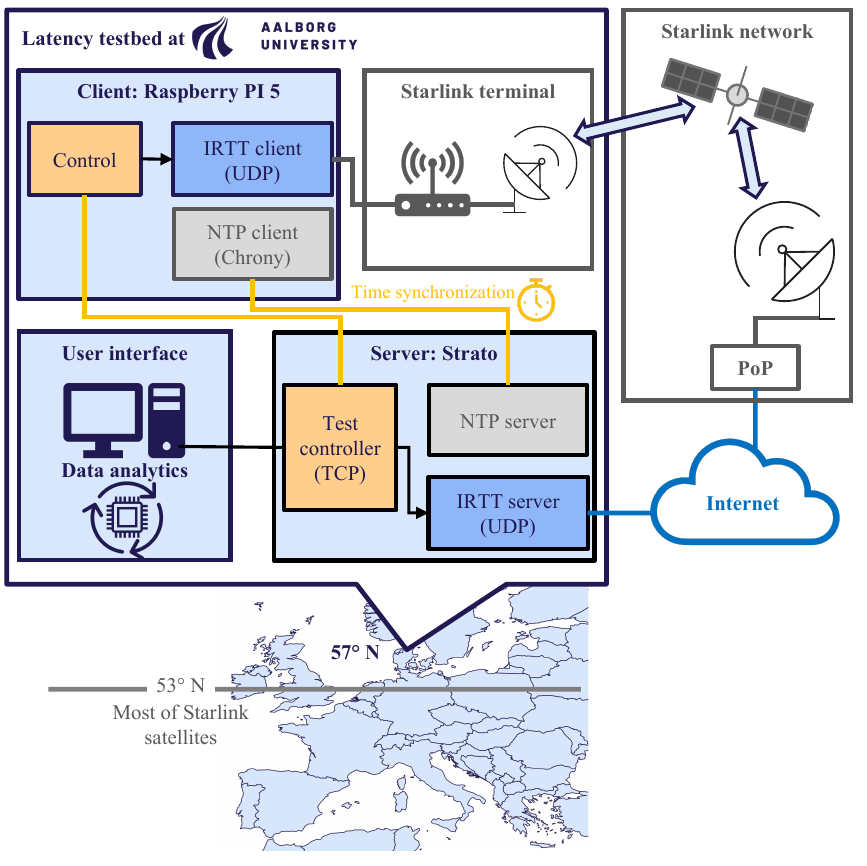}
    \caption{Diagram of the latency testbed at Aalborg University.}
    \label{fig:diag}
\end{figure}

\section{Methods}
\subsection{Latency collection testbed and procedure}
We developed and deployed a local testbed to support the proposed framework by measuring \gls{E2E} latency, both one-way and round-trip, over a \gls{LEO} satellite network. As shown in Fig. \ref{fig:diag}, data is transmitted from a local client (i.e., the source) located physically at Aalborg University (57°N) through the satellite network and up to a server (i.e., the sink) hosted at the same physical location. The collocation of client and server enables tight clock synchronization and full control of both endpoints despite the traffic traversing the satellite network and public internet. 
We use \gls{irtt} \cite{irtt}, a UDP-based tool similar to the \gls{ICMP} ping but designed for high-rate one-way delay measurement at the application layer. Unlike \gls{ICMP}, \gls{irtt} supports non-blocking transmission and does not wait for replies before sending new packets. The client and server clocks are synchronized using Chrony, which implements \gls{NTP} to enable one-way latency estimation with sub-millisecond precision on local area networks \cite{Chrony}.
The probe traffic generated by \gls{irtt} carries only minimal transport-layer overhead and does not emulate higher-layer protocol behavior. As such, the measured delays primarily reflect latency introduced by the network path itself, rather than application-specific processing, and can therefore be interpreted as a lower bound on the latency experienced by real applications.

Latency measurements are collected by transmitting probe packets to record three \gls{E2E} latency metrics: 1) the \gls{UL}, i.e., one-way latency from client to server; 2) the \gls{DL}, i.e., one-way latency from server to client; and 3) the \gls{RTT}, i.e., the latency from client to server and back. Specifically, let $L_{\text{UL}}(t)$, $L_{\text{DL}}(t)$, and $L_{\text{RTT}}(t)$ be the \gls{UL}, \gls{DL}, and \gls{RTT} latency experiences by a packet transmitted at time $t$ by the client. These quantities are related by the decomposition
\begin{equation}
    L_{\text{RTT}}(t) \geq L_{\text{UL}}(t) + L_{\text{DL}}(t).
\end{equation}
Strict equality may not hold due to small local endpoint effects (e.g., host timestamping, OS scheduling) that are not always perfectly accounted for by measurement tools.
Each of these latency components is time-dependent due to satellite dynamics and the status of the transmission queues at the satellites and the ground network elements at the specific time of the packet transmission. Specifically, as shown in Fig.~\ref{fig:diag},  the \gls{E2E} path can be split into a \emph{satellite path} (user terminal → space segment → gateway/\gls{PoP}) and a \emph{terrestrial path} (\gls{PoP} → Internet → Aalborg).
We then write the composition as:
\begin{equation}
    L_i(t) = L_{\mathrm{sat},i}(t) + L_{\mathrm{terr},i}(t),
    \qquad i\in\{\mathrm{UL},\mathrm{DL}\}.
\end{equation}
The satellite path is the highest contributor to latency variations over time and can be further decomposed into propagation, processing, queueing, and handover effects:
\begin{equation}
    L_{\mathrm{sat},i}(t)
    = L_{\mathrm{prop},i}(t) + L_{\mathrm{proc},i}(t) + L_{\mathrm{queue},i}(t) + L_{\mathrm{HO},i}(t).
    \label{eq:sat_decomp}
\end{equation}
Here \(L_{\mathrm{prop},i}\) accounts for propagation along the dish–satellite–gateway hop, \(L_{\mathrm{proc},i}\) for terminal/gateway processing, \(L_{\mathrm{queue},i}\) for queueing in the satellite/gateway path, and \(L_{\mathrm{HO},i}\) for handover-induced excess delay near the 15-second boundaries (often spilling into the subsequent seconds).
In contrast, the terrestrial terms represent the \gls{PoP}-to-Internet-to-Aalborg segment. Based on prior measurements showing that terrestrial Internet paths exhibit relatively stable latency~\cite{davisson2021reassessing, bhat2025constancy}, we assume this component to be approximately time-invariant.

Probe packets were transmitted at 2~ms intervals, corresponding to a sampling rate of 500~Hz. This rate provided the highest temporal resolution supported by the system without inducing significant observer effects. When configured at higher probing rates, we observed increased packet loss and irregular sampling intervals,  indicating that intermediate buffers or scheduling constraints were exceeded.
To ensure stable and reproducible measurements, all subsequent analysis was therefore based on the 500~Hz probing rate, which yields high-resolution insight into the short-term latency dynamics of the satellite link. Prior measurement studies have shown that the handover-related term $L_{\text{HO},i}(t)$ dominates latency near period boundaries; in this work, we explicitly quantify the duration and magnitude of these boundary regions. While the terms $L_{\text{prop,i}}(t) + L_{\mathrm{proc},i}(t) + L_{\text{queue,i}}(t)$ cannot be separated, we investigate their combined impact, striving to characterize how these operate on average and within periods.

The proposed latency characterization framework is provider-agnostic in that it operates solely on end-to-end latency samples and does not rely on satellite identifiers, elevation angles, or routing information. As such, it can be applied to any \gls{LEO} system. The measured latency naturally depends on factors such as satellite geometry, gateway selection, time of day, and network load. To mitigate these effects, we collect datasets across different days and months and verify that the statistical structure found in our analysis remains consistent across traces. Naturally, with time-series measurement, the collected samples are not statistically independent; however, our analysis operates on aggregated intra-period latency distributions rather than individual samples, and therefore does not rely on independence assumptions between consecutive samples.

\section{Results}
\subsection{Period segmentation and summary statistics}
\begin{figure}[t]
    \centering
    \includegraphics[width=1\linewidth]{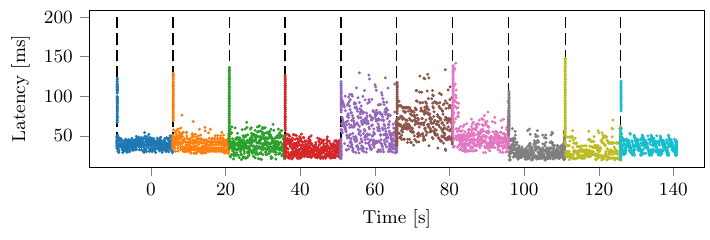}

    \caption{Latency measurements for UL transmission. The beginning of the first period is set as time $t=0$, and the beginning of each period is indicated by a dashed line. Latency measurements for different periods are indicated with different colors.}
    \label{fig:periods}
\end{figure}

As observed in previous studies~\cite{mohan2024multifaceted,pan2024measuring,garcia2024fine}, the \gls{E2E} latency, illustrated in Fig.~\ref{fig:periods} for the \gls{UL}, exhibits clear periodic variations with approximately 15-second intervals. Furthermore, the beginning and end of the 15-second periods are affected by exceptionally high latency. 
Based on the observed periodic behavior, we formulate a statistical data analysis procedure for latency prediction based on time-scale separation that 1) segments the data into 15-second periods and 2) performs a statistical characterization of the latency within individual periods.   

We define $\Delta t=2$~ms to be the period between transmissions of probe packets and $T=15$~s to be the duration of the period in which latency remains stable. Consequently, we collect $S=T/\Delta t=7500$ samples within each 15~s interval. Building on this, we can partition every 15~s period into $S$ discrete time bins, which are indexed by $n$, starting at $n=0$ for a probe packet transmitted at an initial sampling time $t=0$. Next, let $l(n)$ denote the latency at global discrete time bin $n$.
After choosing a reference start $n=0$ for the first period, we map $n=pS+s$ with periods indexed by $p\in\mathcal P=\{0,1,\dots,P-1\}$ and within-period bins by $s\in\mathcal S=\{0,1,\dots,S-1\}$, and define $l_p(s)=l(pS+s)$.
To identify the start of each 15~s period, we apply edge detection to the difference between consecutive samples, $\Delta l(n)=l(n)-l(n-1)$, detecting sharp peaks that correspond to the handover-induced spike between periods. Candidate edges are identified at 2~ms bins where $\Delta l(n)$ exceeds a threshold $\theta$ (robustly chosen). Leveraging the known 15-second periodicity, we enforce a minimum spacing of $T$ between accepted edges. We then map candidate times to phase bins $s=n\bmod S$ and form a histogram over phases,
\begin{equation}\label{eq:phase_hist}
h(s)=\sum_{p\in\mathcal{P}} \mathbf{1}\{\Delta l(pS+s)>\theta\}.
\end{equation}
We locate the most dominant bin, $s_0=\arg\max_s h(s)$. For robustness to binning jitter, we then compute a weighted circular mean over the top k bins in a neighborhood of $s_0$ to obtain a refined start $s^\star$, which we treat as the periodic reference. This reference remains stable across months. The persistent phase alignment suggests the effect is driven by pre-planned, clocked reconfigurations of the network.

\begin{figure}[t]
    \centering
    \includegraphics[width=1\linewidth]{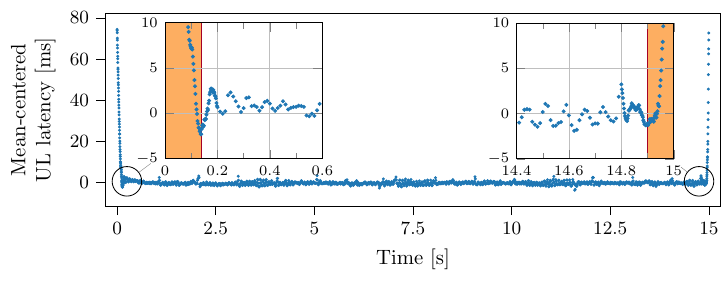}
    \caption{Mean-centered instantaneous uplink latency averaged over multiple 15 second intervals. The red areas, first 140 ms and last 75ms, are the periods affected by a sharp latency increase.}
    \label{fig:delta_w}
\end{figure}

Next, to highlight intra-period dynamics, we remove inter-period variation through mean centering. For $p\in\mathcal P$ and $s\in\mathcal S$, the period mean is
\begin{equation}
 \bar{l}_p = \frac{1}{S}\sum_{s\in\mathcal{S}} l_p(s),   
\end{equation}

and the mean-centered intra-period profile is
\begin{equation}\label{eq:mean_center}
    \hat{l}(s) = \frac{1}{P}\sum_{p\in \mathcal{P}} \bigl(l_p(s) - \bar{l}_p \bigr), 
    \qquad s \in \mathcal{S}.
\end{equation}
By aligning all periods and averaging according to \eqref{eq:mean_center}, yields a representative mean-centered latency profile mapped to the 15-second cycle. 
The results, shown in Fig.~\ref{fig:delta_w}, exhibit a sharp peak at the start of each interval, reaching an average of 74~ms above the mean latency, followed by a rapid decline and a secondary rise near the end of the period. These latency spikes increase jitter and are attributed to reconfiguration events in the Starlink system, such as scheduling changes and handovers. The red regions in Fig.~\ref{fig:delta_w} correspond to the first 140~ms and the last 75~ms of each period; these edge windows are excluded from subsequent analysis to focus on the stable intra-period regime.

\begin{figure}
    \centering
    
    \includegraphics[width=0.8\linewidth]{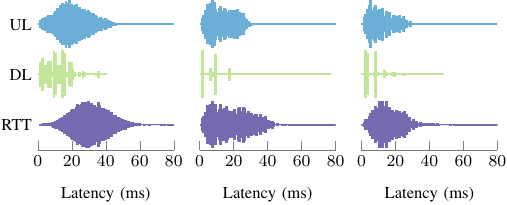}
    \caption{Violin plot showing the intra-period latency for \gls{UL}, \gls{DL}, and \gls{RTT} across three different periods.}
    \label{fig:histograms}
\end{figure}

\begin{figure}
    \centering
    \includegraphics[width=0.8\linewidth]{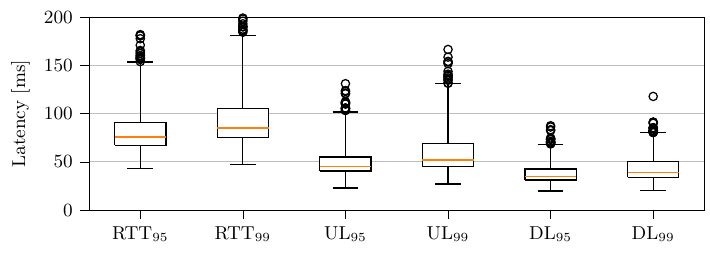}

    \caption{Intra-period latency 95th and 99th quantiles across multiple periods for \gls{UL}, \gls{DL}, and \gls{RTT}.}
    \label{fig:quantiles}
\end{figure}

Although all periods share the same latency spikes at their start and end, the stable intra-period regions exhibit distinct latency distributions. Moreover, \gls{UL}, \gls{DL}, and \gls{RTT} distributions differ considerably across periods. This is illustrated in Fig.~\ref{fig:histograms} for three different periods using a violin plot for \gls{UL}, \gls{DL}, and \gls{RTT}. While the \gls{UL} latency tends to follow a smoother, more continuous distribution, the \gls{DL} latency often appears more concentrated in specific latency values, which is more pronounced for Periods~2 and~3 than for Period~1. This is largely attributed to the bundling of packet transmissions in the \gls{DL} due to queueing at the sink. Building on these differences, we conclude our summary statistics by analysing high quantiles of latency within individual periods. For this, Fig.~\ref{fig:quantiles} shows a box plot with the 95th and 99th quantiles of latency at each individual period for \gls{UL}, \gls{DL}, and \gls{RTT}. The \gls{UL} latency exhibits a broad range and heavier tails than the \gls{DL}, which are also reflected in the \gls{RTT}. In contrast, \gls{DL} latencies remain lower with fewer severe outliers, indicating the \gls{UL}’s dominant role in \gls{RTT} latency variation.

\subsubsection{Model MSE investigation}
The previous analysis shows that Starlink latency exhibits a consistent statistical structure within each 15-second period, which can be leveraged for prediction.
For practical use, however, this predictability must be translated into rapid prediction and classification. Such capabilities are valuable because they allow applications or network functions to adapt transmission strategies early in a period, thereby preserving \gls{QoS} and providing statistical evidence of performance compliance. To maximize the available communication time, predictions should be performed as early as possible within each cycle. In practice, we begin sampling 140~ms after the start of each period—excluding the initial latency spike—and evaluate prediction and classification accuracy as a function of the sampling duration.

Low-latency applications are primarily concerned with upper latency quantiles rather than mean latency. We therefore focus on predicting the 99th percentile of latency within individual periods, considering parametric and non-parametric approaches. Specifically, we investigate the latency profile of \gls{UL} traffic.
In the parametric approach, a distribution must first be selected according to the behavior of the samples to avoid model mismatch. Then, the parameters for the distribution are estimated from the observed data (via maximum likelihood), and the desired quantile is then computed from the fitted distribution.
In our setting, we consider uniform, Gaussian, and Gaussian mixture models (GMM) distributions, defined respectively as follows.
\begin{align}
\mathcal{U}(x \mid a,b) &=
\begin{cases}
\frac{1}{b-a}, & a \leq x \leq b, \\
0, & \text{otherwise},
\end{cases} \\[6pt]
\mathcal{N}(x \mid \mu, \sigma^2) &=
\frac{1}{\sqrt{2\pi\sigma^2}} \exp\!\left(-\frac{(x-\mu)^2}{2\sigma^2}\right), \\[6pt]
\text{GMM}(x) &=
\sum_{k=1}^K \pi_k \, \mathcal{N}(x \mid \mu_k, \sigma_k^2),
\qquad \sum_{k=1}^K \pi_k = 1.
\end{align}
For the non-parametric case, we estimate the 99th percentile directly from the empirical distribution of the samples. The $q$-quantile is given by $F^{-1}(q)$, where $F$ is the CDF of the fitted distribution.
In addition to these parametric and non-parametric approaches, we investigate a tail-oriented modeling strategy based on \gls{EVT}, which is explicitly designed to characterize high-latency events.
When fitting a \gls{GPD} tail model, we follow the peaks-over-threshold approach, where the threshold is implicitly defined by the largest $k$ values  (the \textit{tail fraction}). We select the top-$k$ ($k = 25$) samples in each window to ensure consistent \gls{GPD} modeling across periods. Fixed-quantile-based rules, such as the upper 10\% threshold, can also be applied but are considered inappropriate from a theoretical viewpoint \cite{scarrott2012review}. Because our application requires fitting many short-duration windows, ideally in real time, using a fixed number of exceedances stabilizes \gls{GPD} parameter estimation, mitigates the high variance associated with very high-percentile thresholds, and ensures reduced computational cost.

\begin{figure}
    \centering
    \includegraphics[width=0.8\linewidth]{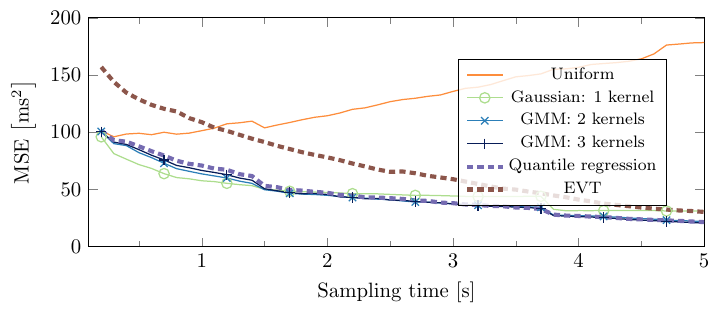}
    \caption{\gls{MSE} for the prediction of the 99-th percentile of \gls{UL} latency as a function of the sampling time.}
    \label{fig:MSE}
\end{figure}

Fig.~\ref{fig:MSE} shows the \gls{MSE} of the 99th percentile prediction as a function of sampling time. We have taken three $\sim$8-hour-long datasets from different days across months to test the generalization of prediction over time.
The parametric estimation with a uniform distribution performs worst overall due to its sensitivity to outliers; as the number of samples increases, extreme values dominate the estimated quantile, reducing accuracy. 
On the other hand, parametric estimation with the Gaussian distribution provides the best short-term accuracy (below 1.5~s), but its error plateaus and even worsens at longer horizons. \gls{EVT} exhibits poor accuracy with very short sampling windows, but improves steadily as more samples are observed.
In contrast, mixture models and non-parametric regression show stable improvement with increasing sample duration, reaching an \gls{MSE} of $\approx 65~\text{ms}^2$ after 1~s and $\approx 21~\text{ms}^2$ after 5~s, corresponding to absolute errors of about 8~ms and 4.5~ms, respectively. However, the improvement rate diminishes beyond the initial sampling phase, indicating that extended sampling yields only marginal gains in predictive accuracy. Similarly, increasing the complexity of mixture models can enhance the representational capacity, but the accuracy gains are marginal and come at the cost of additional computation and fitting time, which limits the practical value.

\subsubsection{Classification and precision-recall}

Once we have identified the distributions that best fit the data, we utilize the temporal dynamics identified in Period segmentation and summary statistics to determine whether a given period can meet application latency requirements. 
A period $p$ is classified as \emph{Good} if at least 99\% of packets meet the latency threshold $l_t$, and as \emph{Degraded} otherwise. Specifically, by defining $\mathbbm{1}\{\cdot\}$ as the indicator function that takes the value of $1$ when the condition between braces is True, we define the underlying class of a period $p$ as
\begin{equation} \label{eq:cases}
\text{Class}(p) = 
\begin{cases}
    \text{Good}, & \frac{1}{S}\sum_{s\in\mathcal{S}} \mathbbm{1}\{l_p(s) \leq l_t\} \geq 0.99, \\[6pt]
    \text{Degraded}, & \text{otherwise}.
\end{cases}
\end{equation}

We apply this classification to define for each period whether they are \emph{Good} or \emph{Degraded}, and investigate how well the fitted models can predict the class. 
The performance of classification with these models is evaluated across different thresholds using \gls{PR} curves, which require a particular probability before designating a prediction as either \emph{Good} or \emph{Degraded}.  
\gls{PR} curves simplify the process of threshold selection by focusing on and quantifying the tradeoff between: 

\begin{itemize}
    \item Precision: Fraction of periods classified as degraded that were correctly classified, which is the ratio of true to total positives. 
    \item Recall: Fraction of degraded periods that were correctly classified, which is the ratio of correctly identified degraded periods to all degraded periods. 
\end{itemize}
These are calculated from the total number of true positives $\mathrm{TP}$, false positives $\mathrm{FP}$, and false negatives $\mathrm{FN}$ as
\begin{equation}
    \text{Precision} = \frac{\mathrm{TP}}{\mathrm{TP} + \mathrm{FP}}, \qquad
    \text{Recall} = \frac{\mathrm{TP}}{\mathrm{TP} + \mathrm{FN}},
\end{equation}
We measure the classification performance using the \gls{AUPRC}, which is well-suited to imbalanced datasets, such as ours, where unstable periods are rarer. A perfect classifier would achieve an \gls{AUPRC} of 1, and lower values indicate a weaker classification performance for Precision and Recall, and require further consideration. 
It is especially interesting to classify a period as accurately and as early as possible, as this would enable the exploitation of good periods, as well as the early adaptation during degraded periods.  The \gls{AUPRC} is computed as
\begin{equation}
    \text{AUPRC} = \sum_{\ell=1}^{L-1} (r_{\ell+1}-r_\ell)\cdot\frac{p_{\ell+1}+p_\ell}{2},
\end{equation}
where $(r_\ell,p_\ell)$ are precision--recall pairs at different threshold levels, of which we have $L$ level pairs. 

\begin{figure}
    \centering
    \includegraphics[width=0.8\linewidth]{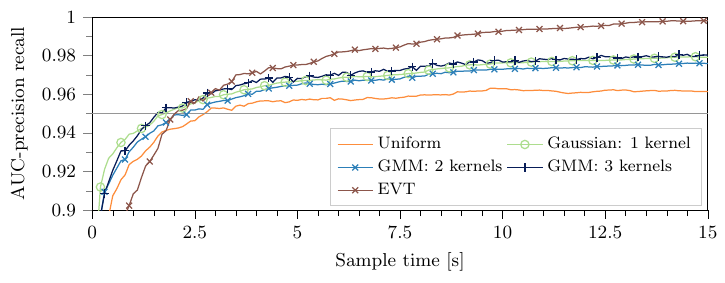}
    \caption{\gls{AUPRC} as a function of the sampling time from the end of the initial latency spike. }
    \label{fig:AUPRC}
\end{figure}

Figure~\ref{fig:AUPRC} shows \gls{AUPRC} as a function of sampling time for a 50 ms threshold ($l_t$) on the 99th latency quantile, which is a typical value for cyber-physical systems, 5G NR satellite access, and the minimal service level for 5G NR  \cite{3GPPCR, 3GPPSR}. 
All models achieve \gls{AUPRC} $\approx 0.85$ within 100~ms. 
In contrast, the \gls{GMM} model with 3 kernels is the fastest to achieve an \gls{AUPRC} of 0.95, taking just 1.6 s. While improvements can be made to the classification performance, such as increasing the number of kernels or discarding outliers, our results show that these lead to negligible improvements of no more than 2\% in the sample time needed to achieve the same \gls{AUPRC} target. Interestingly, \gls{EVT} was slow to outperform on predicting exactly the 99th percentile, but instead exhibits a stronger classification ability. Despite poor initial performance, it outperforms the others after 3.5 seconds and reaches near-perfect classification. While less precise in its predictions, it is more accurate in classifying whether the 99th percentile will exceed or fall below 50ms.

\subsubsection{Discounted service availability}

Finally, we consider a scenario where an intelligent transmission scheduler employs the period-level classification defined in Classification and precision-recall to stop data transmission through the satellite network during periods classified as \emph{degraded}. Such a scheduler is relevant for latency-sensitive applications that must react to predicted violations of their latency budget, as discussed in Introduction.
These transmissions can, therefore, be queued at the client when the application can tolerate short delays. For example, in latency-sensitive applications such as remote operation, commands may be safely delayed or execution temporarily paused until latency returns to acceptable levels, or sent through an additional radio interface if such an interface is available, as considered in multi-connectivity approaches such as \cite{nielsen2017ultra}. In this case, false negatives would lead the scheduler to needlessly stop transmissions during good periods. On the other hand, false positives would not stop data transmission during \emph{degraded} periods, which might lead to degraded performance but no loss of data and, hence, we consider false negatives to be more costly than false positives.

Building on this scenario, we define a holistic score called the \gls{DSA} formulated in equation \ref{eq:dsa}. This is a weighted metric based on the true service availability, defined as the number of good periods $P_\text{Good}$ over a total of $P$ considered periods. Then, the service availability is weighed by 1) the fraction of the available communication time after sampling and 2) the true positive rate, or Recall. Note that the \gls{DSA} metric is upper-bounded by the true service availability (SA), defined as
\begin{equation}
    \mathrm{SA}= \frac{P_\text{Good}}{P}=\frac{\sum_{p=0}^{P-1}\mathbbm{1}\left\{\text{Class}(p)=\text{Good}\right\}}{P}
\end{equation}
equal to 39.6\% for our test dataset.

Furthermore, given that we consider false negatives to be more costly than false positives, the threshold level $\ell$ for classification is defined by setting a maximum \gls{FPR}. The choice of threshold is the first one that can support the maximum false positive rate, which is set to $1$\%, $5$\%, and $10$\% in our experiments.
Then, given a sampling duration $s \cdot \Delta t$  within the $T=15$~s period and a specific \gls{FPR}, the discounted service availability is
\begin{equation}
    \mathrm{DSA}(s, \mathrm{FPR}) = \mathrm{SA} \cdot \dfrac{(T-s\cdot \Delta t)}{T} \cdot \mathrm{TPR}(s \cdot \Delta t)\leq \mathrm{SA}
    \label{eq:dsa}
\end{equation}
where $\text{TPR}(s \cdot \Delta t)$ is the true positive rate achieved with $s \cdot \Delta t$ seconds of sampling. From~\eqref{eq:dsa}, the \gls{DSA} is upper bounded by the true SA because the true positive rate cannot exceed $1$ and the probing duration $s \cdot \Delta t$ is positive and upper bounded by $T$.

\begin{figure}
    \centering
    \includegraphics[width=0.8\linewidth]{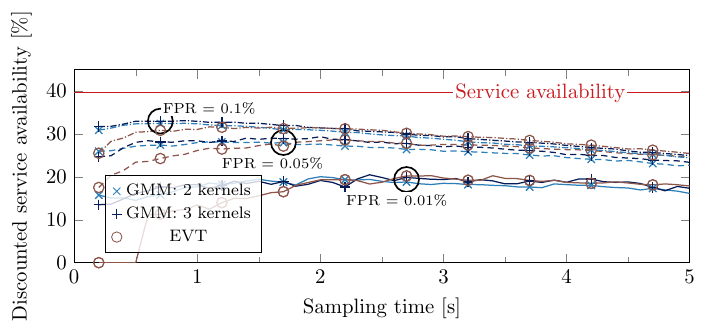}
    \caption{\gls{DSA} as a function of the sampling time from the beginning of the period for a pre-defined maximum \gls{FPR}.}
    \label{fig:DSA}
\end{figure}

Next, we evaluate the \gls{DSA} at different sample times and \glspl{FPR} for both GMM and \gls{EVT} predictors. Figure~\ref{fig:DSA} shows the resulting \gls{DSA} curves for GMM models with 2 and 3 kernels with 3 different defined maximum acceptable \gls{FPR}. 
The graphs indicate that relaxing the \gls{FPR} leads to a higher score for the \gls{DSA} formulation. This follows from stringent requirements, which necessitate more sampling to achieve the desired \gls{FPR} accurately. Consequently, the \gls{DSA} decreases as less time is available for communication. Comparatively, GMM can rapidly achieve good prediction performance, whereas \gls{EVT} requires additional samples; however, this results in less time for using the link, leading to a lower achievable DSA for \gls{EVT}.

Our findings confirm that lightweight statistical models can achieve near-optimal classification performance with only limited sampling. This enables the fine-tuning of the classifier to achieve a specific trade-off between false positives and detection delay, benefiting specific applications. This approach also provides a practical balance between responsiveness and service reliability. Therefore, our methods can be leveraged according to the requirements of specific applications and the capabilities of specific systems, for example, based on the availability of multiple communication interfaces. Specifically, for systems with multi-connectivity or interface-diversity capabilities, the early detection of degraded periods can be used for steering the traffic between interfaces and/or networks, enabling real-time and efficient system adaptation.


\section{Discussion}
\glsresetall

In this article, we presented a statistical framework for period-level characterization, prediction, and classification of end-to-end latency in LEO satellite networks. By applying these methods to Starlink's system, we characterized its deterministic 15-second periodic structure, investigating \gls{UL}, \gls{DL}, and \gls{RTT}. We also showed that the initial and final portions of each cycle consistently exhibit deterministically high latency, attributable to satellite handovers, and thus provide limited predictive value. 
Excluding these boundary regions, we focused on the stable intra-period core and demonstrated that lightweight parametric and non-parametric models can predict and classify latency with high accuracy from only a few samples. Predictive accuracy was quantified using both regression (\gls{MSE}) and classification (\gls{AUPRC}), confirming the feasibility of simple yet effective forecasting approaches.  
These findings highlight the potential of predictive middleware to leverage Starlink’s deterministic latency structure more effectively. 
By anticipating low- and high-latency windows, latency-sensitive applications can proactively adapt through pacing, buffering, quality adjustment, or safe-mode transitions, thereby improving quality of service and user experience.
Our results provide the statistical foundation for such middleware, which can operate passively in the background to enable application-aware adaptation over LEO satellite networks, thereby achieving optimal performance while accounting for the inherent uncertainty of satellite links. 

Future work could investigate how specific transport and application protocols interact with the characterized latency dynamics, providing further insights for protocol design in LEO satellite networks. Additionally, extending this characterization framework to other LEO constellations would validate its generalizability across different satellite systems.


\begin{thebibliography}{23}
\ifx \bisbn   \undefined \def \bisbn  #1{ISBN #1}\fi
\ifx \binits  \undefined \def \binits#1{#1}\fi
\ifx \bauthor  \undefined \def \bauthor#1{#1}\fi
\ifx \batitle  \undefined \def \batitle#1{#1}\fi
\ifx \bjtitle  \undefined \def \bjtitle#1{#1}\fi
\ifx \bvolume  \undefined \def \bvolume#1{\textbf{#1}}\fi
\ifx \byear  \undefined \def \byear#1{#1}\fi
\ifx \bissue  \undefined \def \bissue#1{#1}\fi
\ifx \bfpage  \undefined \def \bfpage#1{#1}\fi
\ifx \blpage  \undefined \def \blpage #1{#1}\fi
\ifx \burl  \undefined \def \burl#1{\textsf{#1}}\fi
\ifx \doiurl  \undefined \def \doiurl#1{\url{https://doi.org/#1}}\fi
\ifx \betal  \undefined \def \betal{\textit{et al.}}\fi
\ifx \binstitute  \undefined \def \binstitute#1{#1}\fi
\ifx \binstitutionaled  \undefined \def \binstitutionaled#1{#1}\fi
\ifx \bctitle  \undefined \def \bctitle#1{#1}\fi
\ifx \beditor  \undefined \def \beditor#1{#1}\fi
\ifx \bpublisher  \undefined \def \bpublisher#1{#1}\fi
\ifx \bbtitle  \undefined \def \bbtitle#1{#1}\fi
\ifx \bedition  \undefined \def \bedition#1{#1}\fi
\ifx \bseriesno  \undefined \def \bseriesno#1{#1}\fi
\ifx \blocation  \undefined \def \blocation#1{#1}\fi
\ifx \bsertitle  \undefined \def \bsertitle#1{#1}\fi
\ifx \bsnm \undefined \def \bsnm#1{#1}\fi
\ifx \bsuffix \undefined \def \bsuffix#1{#1}\fi
\ifx \bparticle \undefined \def \bparticle#1{#1}\fi
\ifx \barticle \undefined \def \barticle#1{#1}\fi
\bibcommenthead
\ifx \bconfdate \undefined \def \bconfdate #1{#1}\fi
\ifx \botherref \undefined \def \botherref #1{#1}\fi
\ifx \url \undefined \def \url#1{\textsf{#1}}\fi
\ifx \bchapter \undefined \def \bchapter#1{#1}\fi
\ifx \bbook \undefined \def \bbook#1{#1}\fi
\ifx \bcomment \undefined \def \bcomment#1{#1}\fi
\ifx \oauthor \undefined \def \oauthor#1{#1}\fi
\ifx \citeauthoryear \undefined \def \citeauthoryear#1{#1}\fi
\ifx \endbibitem  \undefined \def \endbibitem {}\fi
\ifx \bconflocation  \undefined \def \bconflocation#1{#1}\fi
\ifx \arxivurl  \undefined \def \arxivurl#1{\textsf{#1}}\fi
\csname PreBibitemsHook\endcsname

\bibitem[\protect\citeauthoryear{L{\'o}pez et~al.}{2023}]{lopez2023connecting}
\begin{bchapter}
\bauthor{\bsnm{L{\'o}pez}, \binits{M.}},
\bauthor{\bsnm{Damsgaard}, \binits{S.B.}},
\bauthor{\bsnm{Rodr{\'\i}guez}, \binits{I.}},
\bauthor{\bsnm{Mogensen}, \binits{P.}}:
\bctitle{Connecting rural areas: an empirical assessment of 5g terrestrial-leo satellite multi-connectivity}.
In: \bbtitle{2023 IEEE 97th Vehicular Technology Conference (VTC2023-Spring)},
pp. \bfpage{1}--\blpage{5}
(\byear{2023}).
\bcomment{IEEE}
\end{bchapter}
\endbibitem

\bibitem[\protect\citeauthoryear{Zhao and Pan}{2024}]{zhao2024low}
\begin{bchapter}
\bauthor{\bsnm{Zhao}, \binits{J.}},
\bauthor{\bsnm{Pan}, \binits{J.}}:
\bctitle{Low-latency live video streaming over a low-earth-orbit satellite network with dash}.
In: \bbtitle{Proceedings of the 15th ACM Multimedia Systems Conference},
pp. \bfpage{109}--\blpage{120}
(\byear{2024})
\end{bchapter}
\endbibitem

\bibitem[\protect\citeauthoryear{Ma et~al.}{2023}]{ma2023network}
\begin{bchapter}
\bauthor{\bsnm{Ma}, \binits{S.}},
\bauthor{\bsnm{Chou}, \binits{Y.C.}},
\bauthor{\bsnm{Zhao}, \binits{H.}},
\bauthor{\bsnm{Chen}, \binits{L.}},
\bauthor{\bsnm{Ma}, \binits{X.}},
\bauthor{\bsnm{Liu}, \binits{J.}}:
\bctitle{Network characteristics of leo satellite constellations: A starlink-based measurement from end users}.
In: \bbtitle{IEEE INFOCOM 2023-IEEE Conference on Computer Communications},
pp. \bfpage{1}--\blpage{10}
(\byear{2023}).
\bcomment{IEEE}
\end{bchapter}
\endbibitem

\bibitem[\protect\citeauthoryear{Mohan et~al.}{2024}]{mohan2024multifaceted}
\begin{bchapter}
\bauthor{\bsnm{Mohan}, \binits{N.}},
\bauthor{\bsnm{Ferguson}, \binits{A.E.}},
\bauthor{\bsnm{Cech}, \binits{H.}},
\bauthor{\bsnm{Bose}, \binits{R.}},
\bauthor{\bsnm{Renatin}, \binits{P.R.}},
\bauthor{\bsnm{Marina}, \binits{M.K.}},
\bauthor{\bsnm{Ott}, \binits{J.}}:
\bctitle{A multifaceted look at starlink performance}.
In: \bbtitle{Proceedings of the ACM on Web Conference 2024},
pp. \bfpage{2723}--\blpage{2734}
(\byear{2024})
\end{bchapter}
\endbibitem

\bibitem[\protect\citeauthoryear{Pan et~al.}{2024}]{pan2024measuring}
\begin{bchapter}
\bauthor{\bsnm{Pan}, \binits{J.}},
\bauthor{\bsnm{Zhao}, \binits{J.}},
\bauthor{\bsnm{Cai}, \binits{L.}}:
\bctitle{Measuring the satellite links of a leo network}.
In: \bbtitle{IEEE International Conference on Communications}
(\byear{2024})
\end{bchapter}
\endbibitem

\bibitem[\protect\citeauthoryear{Garcia et~al.}{2024}]{garcia2024fine}
\begin{bchapter}
\bauthor{\bsnm{Garcia}, \binits{J.}},
\bauthor{\bsnm{Sundberg}, \binits{S.}},
\bauthor{\bsnm{Brunstrom}, \binits{A.}}:
\bctitle{Fine-grained starlink throughput variation examined with state-transition modeling}.
In: \bbtitle{2024 19th Wireless On-Demand Network Systems and Services Conference (WONS)},
pp. \bfpage{69}--\blpage{76}
(\byear{2024}).
\bcomment{IEEE}
\end{bchapter}
\endbibitem

\bibitem[\protect\citeauthoryear{Husseyn et~al.}{2025}]{inflation}
\begin{barticle}
\bauthor{\bsnm{Husseyn}, \binits{D.}},
\bauthor{\bsnm{Saranya}, \binits{D.G.}},
\bauthor{\bsnm{Babu}, \binits{D.K.}},
\bauthor{\bsnm{Kishore}, \binits{D.D.}},
\bauthor{\bsnm{Kiruthikadevi}, \binits{K.}}:
\batitle{Characterizing latency inflation in mobile and satellite internet connections}.
\bjtitle{Journal of Internet Services and Information Security}
\bvolume{15},
\bfpage{573}--\blpage{586}
(\byear{2025})
\doiurl{10.58346/JISIS.2025.I3.039}
\end{barticle}
\endbibitem

\bibitem[\protect\citeauthoryear{Tiwari et~al.}{2023}]{tiwari2023t3p}
\begin{botherref}
\oauthor{\bsnm{Tiwari}, \binits{S.}},
\oauthor{\bsnm{Bhushan}, \binits{S.}},
\oauthor{\bsnm{Taneja}, \binits{A.}},
\oauthor{\bsnm{Kassem}, \binits{M.}},
\oauthor{\bsnm{Luo}, \binits{C.}},
\oauthor{\bsnm{Zhou}, \binits{C.}},
\oauthor{\bsnm{He}, \binits{Z.}},
\oauthor{\bsnm{Raman}, \binits{A.}},
\oauthor{\bsnm{Sastry}, \binits{N.}},
\oauthor{\bsnm{Qiu}, \binits{L.}}, et al.:
T3p: Demystifying low-earth orbit satellite broadband.
arXiv preprint arXiv:2310.11835
(2023)
\end{botherref}
\endbibitem

\bibitem[\protect\citeauthoryear{Rojas et~al.}{2025}]{rojas2025latency}
\begin{bchapter}
\bauthor{\bsnm{Rojas}, \binits{C.}},
\bauthor{\bsnm{Fraire}, \binits{J.A.}},
\bauthor{\bsnm{Patrone}, \binits{F.}},
\bauthor{\bsnm{Marchese}, \binits{M.}}:
\bctitle{On the latency trade-off between space and terrestrial clouds in non-terrestrial networks}.
In: \bbtitle{2025 12th Advanced Satellite Multimedia Systems Conference and the 18th Signal Processing for Space Communications Workshop (ASMS/SPSC)},
pp. \bfpage{1}--\blpage{8}
(\byear{2025}).
\bcomment{IEEE}
\end{bchapter}
\endbibitem

\bibitem[\protect\citeauthoryear{van’t Hof et~al.}{2019}]{van2019low}
\begin{bchapter}
\bauthor{\bsnm{Hof}, \binits{J.}},
\bauthor{\bsnm{Karunanithi}, \binits{V.}},
\bauthor{\bsnm{Speretta}, \binits{S.}},
\bauthor{\bsnm{Verhoeven}, \binits{C.}},
\bauthor{\bsnm{McCune}, \binits{E.}}:
\bctitle{Low latency iot/m2m using nano-satellites}.
In: \bbtitle{70th International Astronautical Congress (IAC), Washington DC, United States},
pp. \bfpage{21}--\blpage{25}
(\byear{2019})
\end{bchapter}
\endbibitem

\bibitem[\protect\citeauthoryear{Liu et~al.}{2025}]{liu2025vivisecting}
\begin{barticle}
\bauthor{\bsnm{Liu}, \binits{Z.}},
\bauthor{\bsnm{Reidys}, \binits{F.-X.G.}},
\bauthor{\bsnm{Tanveer}, \binits{S.}},
\bauthor{\bsnm{Vasisht}, \binits{D.}}:
\batitle{Vivisecting starlink throughput: Measurement and prediction}.
\bjtitle{Proceedings of the ACM on Networking}
\bvolume{3}(\bissue{CoNEXT4}),
\bfpage{1}--\blpage{23}
(\byear{2025})
\end{barticle}
\endbibitem

\bibitem[\protect\citeauthoryear{Zhu and Lu}{2024}]{zhu2024latency}
\begin{barticle}
\bauthor{\bsnm{Zhu}, \binits{Y.}},
\bauthor{\bsnm{Lu}, \binits{D.}}:
\batitle{Latency guaranteed joint optimal traffic intelligent scheduling in large-scale satellite networks}.
\bjtitle{Computer Networks}
\bvolume{242},
\bfpage{110236}
(\byear{2024})
\end{barticle}
\endbibitem

\bibitem[\protect\citeauthoryear{L{\'o}pez et~al.}{2022}]{lopez202212}
\begin{bchapter}
\bauthor{\bsnm{L{\'o}pez}, \binits{M.}},
\bauthor{\bsnm{Damsgaard}, \binits{S.B.}},
\bauthor{\bsnm{Rodr{\'\i}guez}, \binits{I.}},
\bauthor{\bsnm{Mogensen}, \binits{P.}}:
\bctitle{An empirical analysis of multi-connectivity between 5g terrestrial and leo satellite networks}.
In: \bbtitle{Proc. IEEE Globecom Workshops (GC Wkshps)},
pp. \bfpage{1115}--\blpage{1120}
(\byear{2022})
\end{bchapter}
\endbibitem

\bibitem[\protect\citeauthoryear{Garcia et~al.}{2023}]{garcia2023multi}
\begin{bchapter}
\bauthor{\bsnm{Garcia}, \binits{J.}},
\bauthor{\bsnm{Sundberg}, \binits{S.}},
\bauthor{\bsnm{Caso}, \binits{G.}},
\bauthor{\bsnm{Brunstrom}, \binits{A.}}:
\bctitle{Multi-timescale evaluation of starlink throughput}.
In: \bbtitle{Proceedings of the 1st ACM Workshop on LEO Networking and Communication},
pp. \bfpage{31}--\blpage{36}
(\byear{2023})
\end{bchapter}
\endbibitem

\bibitem[\protect\citeauthoryear{Lai et~al.}{2022}]{lai2022spacertc}
\begin{bchapter}
\bauthor{\bsnm{Lai}, \binits{Z.}},
\bauthor{\bsnm{Liu}, \binits{W.}},
\bauthor{\bsnm{Wu}, \binits{Q.}},
\bauthor{\bsnm{Li}, \binits{H.}},
\bauthor{\bsnm{Xu}, \binits{J.}},
\bauthor{\bsnm{Wu}, \binits{J.}}:
\bctitle{Spacertc: Unleashing the low-latency potential of mega-constellations for real-time communications}.
In: \bbtitle{IEEE INFOCOM 2022-IEEE Conference on Computer Communications},
pp. \bfpage{1339}--\blpage{1348}
(\byear{2022}).
\bcomment{IEEE}
\end{bchapter}
\endbibitem

\bibitem[\protect\citeauthoryear{Heist}{2016}]{irtt}
\begin{botherref}
\oauthor{\bsnm{Heist}, \binits{P.}}:
{IRTT} (Isochronous Round-Trip Tester).
Accessed: 2025-09-09
(2016).
\url{https://github.com/heistp/irtt}
\end{botherref}
\endbibitem

\bibitem[\protect\citeauthoryear{{Chrony Project}}{2024}]{Chrony}
\begin{botherref}
\oauthor{\bsnm{{Chrony Project}}}:
Configuration examples and accuracy.
Accessed: 2025-09-09
(2024).
\url{https://chrony-project.org/examples.html}
\end{botherref}
\endbibitem

\bibitem[\protect\citeauthoryear{Davisson et~al.}{2021}]{davisson2021reassessing}
\begin{bchapter}
\bauthor{\bsnm{Davisson}, \binits{L.}},
\bauthor{\bsnm{Jakovleski}, \binits{J.}},
\bauthor{\bsnm{Ngo}, \binits{N.}},
\bauthor{\bsnm{Pham}, \binits{C.}},
\bauthor{\bsnm{Sommers}, \binits{J.}}:
\bctitle{Reassessing the constancy of end-to-end internet latency}.
In: \bbtitle{IFIP Network Traffic Measurement and Analysis Conference}
(\byear{2021})
\end{bchapter}
\endbibitem

\bibitem[\protect\citeauthoryear{Bhat et~al.}{2025}]{bhat2025constancy}
\begin{bchapter}
\bauthor{\bsnm{Bhat}, \binits{A.}},
\bauthor{\bsnm{Ganatra}, \binits{V.}},
\bauthor{\bsnm{Shaha}, \binits{A.}},
\bauthor{\bsnm{Chandrasekaran}, \binits{B.}},
\bauthor{\bsnm{Naik}, \binits{V.}}:
\bctitle{On the constancy of latency at the internet's edge}.
In: \bbtitle{2025 9th Network Traffic Measurement and Analysis Conference (TMA)},
pp. \bfpage{1}--\blpage{10}
(\byear{2025}).
\bcomment{IEEE}
\end{bchapter}
\endbibitem

\bibitem[\protect\citeauthoryear{Scarrott and MacDonald}{2012}]{scarrott2012review}
\begin{barticle}
\bauthor{\bsnm{Scarrott}, \binits{C.}},
\bauthor{\bsnm{MacDonald}, \binits{A.}}:
\batitle{A review of extreme value threshold estimation and uncertainty quantification}.
\bjtitle{REVSTAT-Statistical journal}
\bvolume{10}(\bissue{1}),
\bfpage{33}--\blpage{60}
(\byear{2012})
\end{barticle}
\endbibitem

\bibitem[\protect\citeauthoryear{{3rd Generation Partnership Project (3GPP)}}{2024}]{3GPPCR}
\begin{botherref}
\oauthor{\bsnm{{3rd Generation Partnership Project (3GPP)}}}:
Service requirements for cyber-physical control applications in vertical domains.
3GPP TS 22.104, V18.4.0.
Accessed: 2025-09-09
(2024).
\url{https://www.3gpp.org/ftp/Specs/archive/22_series/22.104/22104-h40.zip}
\end{botherref}
\endbibitem

\bibitem[\protect\citeauthoryear{{3rd Generation Partnership Project (3GPP)}}{2025}]{3GPPSR}
\begin{botherref}
\oauthor{\bsnm{{3rd Generation Partnership Project (3GPP)}}}:
Service requirements for the 5G system; Stage 1.
3GPP TS 22.261, V20.2.0.
Accessed: 2025-09-09
(2025).
\url{https://www.3gpp.org/ftp/Specs/archive/22_series/22.261/22261-i20.zip}
\end{botherref}
\endbibitem

\bibitem[\protect\citeauthoryear{Nielsen et~al.}{2017}]{nielsen2017ultra}
\begin{barticle}
\bauthor{\bsnm{Nielsen}, \binits{J.J.}},
\bauthor{\bsnm{Liu}, \binits{R.}},
\bauthor{\bsnm{Popovski}, \binits{P.}}:
\batitle{Ultra-reliable low latency communication using interface diversity}.
\bjtitle{IEEE Transactions on Communications}
\bvolume{66}(\bissue{3}),
\bfpage{1322}--\blpage{1334}
(\byear{2017})
\end{barticle}
\endbibitem

\end{thebibliography}
\end{document}